\documentclass{article}
\usepackage{epsf}
\usepackage{amssymb}
\newcommand{\bookfig}[5]{
\begin{figure}\centering\fbox{\epsfysize=#5cm \epsfbox{#1}}
\caption[#2]{\small #4}\label{#3}
\end{figure}
}

\newcommand{\be}{\begin{equation}}
\newcommand{\ee}{\end{equation}}
\newcommand{\bea}{\begin{eqnarray}}
\newcommand{\eea}{\end{eqnarray}}
\newcommand{\beas}{\begin{eqnarray*}}
\newcommand{\eeas}{\end{eqnarray*}}

\def\build#1_#2^#3{\mathrel{
\mathop{\kern 0pt#1}\limits_{#2}^{#3}}}

\newtheorem{theorem}{Theorem}

\newtheorem{prop}[theorem]{Proposition}

\font\tenbb=msbm10 \font\sevenbb=msbm7 \font\fivebb=msbm5
\newfam\bbfam
\textfont\bbfam=\tenbb \scriptfont\bbfam=\sevenbb
\scriptscriptfont\bbfam=\fivebb

\def\d{\delta}

\def\ve{\varepsilon}

\font\tenbb=msbm10 \font\sevenbb=msbm7 \font\fivebb=msbm5
\newfam\bbfam
\textfont\bbfam=\tenbb \scriptfont\bbfam=\sevenbb
\scriptscriptfont\bbfam=\fivebb

\catcode`\@=11
\def\displaylinesno #1{\displ@y\halign{
\hbox to\displaywidth{$\@lign\hfil\displaystyle##\hfil$}&
\llap{$##$}\crcr#1\crcr}}

\def\ldisplaylinesno #1{\displ@y\halign{
\hbox to\displaywidth{$\@lign\hfil\displaystyle##\hfil$}&
\kern-\displaywidth\rlap{$##$} \tabskip\displaywidth\crcr#1\crcr}}
\catcode`\@=12

\def\build#1_#2^#3{\mathrel{
\mathop{\kern 0pt#1}\limits_{#2}^{#3}}}

\def\vvp{\;\raisebox{-1mm}{\epsfysize=4mm\epsfbox{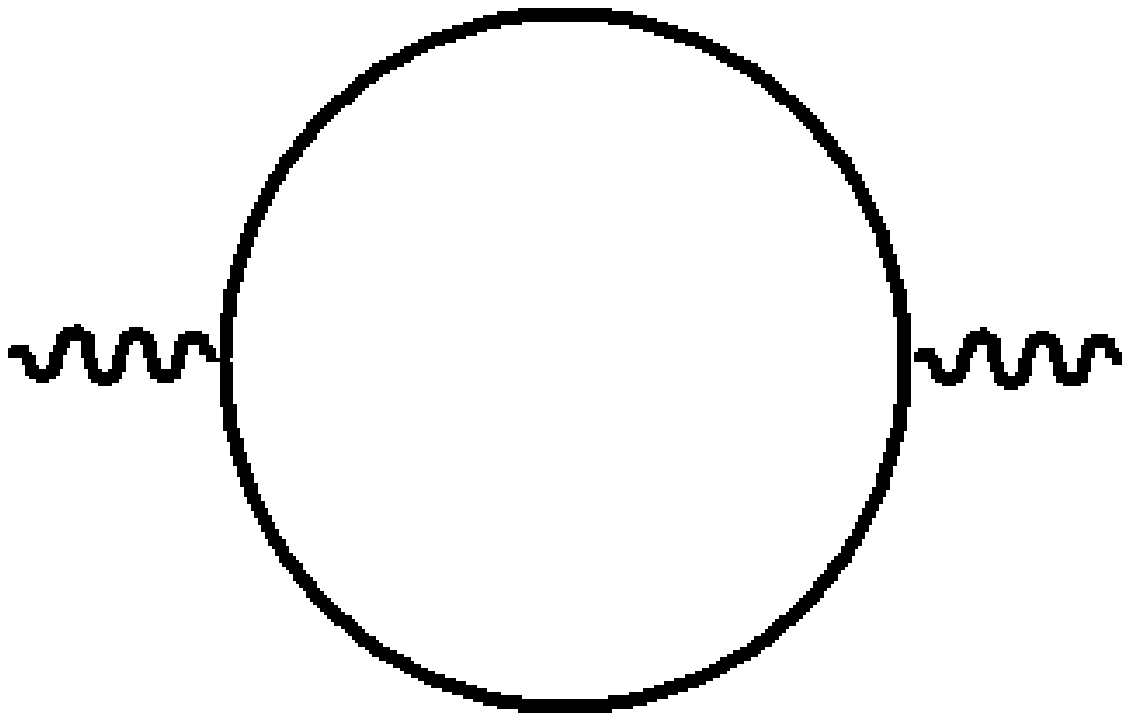}}\;}

\begin{document}
\title{\bf New mathematical structures in renormalizable quantum field theories\footnote{Invited Contribution
to the January '03 special edition of Annals of Physics.}}
\author{Dirk Kreimer\thanks{kreimer@ihes.fr}
} \maketitle \vspace{-24em}\begin{flushright} {\small
ESI-1236\\
BUCMP/02-05\\
hep-th/0211136 }
\end{flushright}\vspace{20em}

\begin{abstract}
Computations in renormalizable perturbative quantum field theories
reveal mathematical structures which go way beyond the formal
structure which is usually taken as underlying quantum field
theory. We review these new structures and the role they can play
in future developments.
\end{abstract}
{\footnotesize {\bf PACS classification:} 11.10Gh 11.15Bt 11.10Jj}
\section*{Introduction}
Quantum field theory is a venerable subject by now as the sole
means  providing us on a daily basis with insights into the laws
of nature  in the high energy laboratories around the world. Its
most spectacular successes are in the perturbative regime, where it
provides for much celebrated coincidence between radiative
correction calculations and experiment. Similarly successful is
its Euclidean counterpart in the realms of statistical physics
\cite{Zinn-Justin}.

While demanding in their technical details, the computational
praxis of these calculations has essentially remained the same
since loop calculations started in earnest several decades ago:\\
- recursively, construct local counterterms so as to make any term
in the perturbative expansion finite,\\ -  find finite
renormalizations such that the Ward--Takahashi- and
Slavnov--Taylor identities are respected order by order,\\ - and
finally, calculate as much as you can.

The Standard Model fares notoriously well when subjected to this
program, and in particular in its radiative correction sector it
allows for an indirect look at inaccessible high energies, with
results which so far do not support any deviation from that model
in any conclusive manner.

It is well understood how to set up such calculations in
accordance with the requirements of quantized local gauge
theories. Here and there one or the other technicality still
demands clarification (see \cite{non-Aoki} for an example), but by
now the dedicated group of practitioners of quantum field theory
has the technical challenges implied by locality, causality and
internal symmetries well under control.

The surprises and challenges for those practitioners of quantum
field theory come from a rather unexpected direction: there is, in
this praxis of computational quantum field theory seemingly
overloaded by technicalities, a clear sign of deeper mathematical
structure underlying quantum field theory which starts to emerge
when one investigates the structure of higher order terms in the
celebrated loop expansion.

For me, the two big surprises hidden in high loop order
calculations are\\ - the number theoretic content of QFT and\\
- the Lie algebra of Feynman graphs overlooked for half a century.

They both, I will argue, combine towards pointing to a connection
of quantum physics to number theory which, to my mind, must
 be understood before we have any hope of deciphering the
message of physics at small distances in any meaningful way.

Both surprises are  typical perturbative phenomena. Both, I
believe, tell us something about the exact theory which none of
the so-called rigorous approaches to quantum field theory seems
yet to be able to reveal. Indeed, it seems to be a notorious
property of perturbation theory that this sum of the parts is
larger than the whole, in the sense that quite often the
perturbative expansion is more revealing even in circumstances
where an exact solution is available \cite{Marino}.

In this sense, our venerable subject of QFT is still rather
juvenile: we are only at the beginning of getting an idea about
the transcendental nature of the numbers and special functions in
its realm. Even more baffling, the Lie algebras underlying Feynman
graphs are at this moment poorly investigated whilst apparently
very rich in structure: the question to what degree the secrets of
the physics of the very small lie hidden in their representation
theory we have only very recently learned to ask.

In this overview we want to describe mathematical structures in
renormalizable quantum field theories as they were discovered
recently. We focus on renormalizable theories in four dimensions
of spacetime and their perturbative expansion in terms of Feynman
graphs, with emphasis given to possible non-perturbative aspects.

We will review recent results concerning the Hopf and Lie algebra
structures in such theories first. From there, we will connect
them to various branches in mathematics, foremost among them number theory,
and also to selected aspects of noncommutative geometry.

We also will present some new results, with a detailed derivation
given elsewhere, and will continuously point out open questions
and perspectives.

Almost all the material presented stems from practical experience
with the calculation of Feynman graphs. Indeed, our viewpoint is
quite similar to that of 't Hooft and Veltman's DIAGRAMMAR
\cite{diag}: in the absence of a truly rigorous derivation
of Feynman rules, let us take Feynman diagrams as the starting
point and try to understand their structure. It is most amazing to
what extent combinatorial and graph-theoretic structures already
prescribe the properties which are usually celebrated as the
triumph of the axiomatic underpinning of QFT. It is most
gratifying indeed to see locality emerge just from basic
combinatorial properties of Lie and Hopf algebras of graphs, and
even more gratifying to my mind to see the close relation to
$\zeta$-functions already emerge at a combinatorial level. A
further treat along these lines is the emergence of the
renormalization group from the consideration of one-parameter
groups of automorphisms of this Hopf algebra, and the final
culmination of these structures in the Riemann--Hilbert problem
and its connection to renormalization theory \cite{RHI,RHII} .

None of this is in conflict with the standard lore on QFT as
developed in the 1970's. What is at stake though is the question
of how fundamental this textbook approach is. The hints are growing
that there is a deeper level possible in the understanding of QFT
and that the axiomatics of QFT are, possibly, corollaries of yet
to be discovered mathematical structures, structures which all
celebrate the fundamental role played by locality and its
consequences in the elimination of short-distance singularities.
The emergence of beautiful structures in the concepts of
renormalization theory only emphasizes the importance of the
groundwork of the fathers of renormalization theory for future
progress with QFT.

In section one we summarize the basic notions of perturbative
quantum field theory using the pre-Lie structure of graph
insertions. This allows us to derive forest formulas for
renormalization in a rather succinct manner. The basic route
towards a perturbative quantum field theory from this viewpoint
is: \begin{itemize} \item decide what the field content is of your
theory, appropriately specifying quantum numbers (spin, mass,
flavor, color and such) of fields, restricting interactions so as
to obtain a renormalizable theory; \item consider all 1PI graphs
you can construct with edges corresponding to free-field
covariances and vertices for local interactions and realize that
they allow for a pre-Lie algebra of graph insertions.
Antisymmetrize this pre-Lie product to get a Lie algebra of graph
insertions and consider the Hopf algebra which is dual to the
enveloping algebra of this Lie algebra \cite{RHI,InsElim}; \item
realize that the coproduct and antipode of this Hopf algebra give
rise to the forest formula which generates local counterterms upon
introducing a Rota--Baxter map, a renormalization scheme in 
physicists' parlance \cite{DK1,Chen};
\item use the Hochschild cohomology of this Hopf algebra to prove
finiteness of renormalized graphs and to show that you can absorb
singularities in local counterterms \cite{RHI,DK1,dennis}; \item
use the full Hopf algebra of graphs (which has the structure of a
semi-direct product of superficially divergent graphs with
convergent ones) to obtain the finite renormalization needed to
satisfy the requirements of quantized gauge symmetries
\cite{RHI,Chen}.\end{itemize} This structure underlies any of
the approaches to perturbative quantum field theory, and wether we do
$x$-space methods or momentum space methods is essentially a matter
of taste and practical consideration, which often favor momentum
space integrations. The beautiful number-theoretic aspects of
perturbative quantum field theory would still lay dormant were it
not for momentum space integration methods which allow us to gather
evidence at three loops and far beyond
\cite{Book,BKold1,BKold2,BGK,BK1,BK2,BK4}.

Immediate questions which arise from this viewpoint, partially
answered in the literature,  are the classification of
renormalization schemes in terms of Rota--Baxter algebras
\cite{DK1,Chen}, an exploration of the amazing  connection to the
Riemann--Hilbert problem which emerges in the context where the
Rota--Baxter map is a minimal subtraction using an analytic
regularization parameter \cite{RHI,RHII}, and the study of
homomorphisms of the Lie group -associated to the Lie algebra of
graphs-  to diffeomorphism groups of physical parameters, which
establishes the perturbative renormalization group via its
one-parameter group of automorphisms \cite{RHII}. A short review
of these results finishes section one.

In section two we consider perspectives and work in progress
emerging from the results reported in section one. Our main point
is the discussion of a connection between Euler products and
quantum field theory. We start with the Riemann $\zeta$-function
and derive it as a solution to a Dyson--Schwinger equation. This
is only meant as motivation to reverse the process and to look for
Euler products in quantum field theory in general. These products
are obtained using a symmetrized product of graph insertions
induced in the Hopf algebra by the pre-Lie structure in the dual.
We discuss the structure of a formal solution to a
Dyson--Schwinger equation in terms of Euler products of primitive
graphs. In particular, we find that questions about gauge
symmetries are intimately connected with such factorizations. This
raises one central question: how do such factorizations fare under
evaluation by the Feynman rules? Is the evaluation of a product
related to the product of the evaluations? Before we can address
this question in a meaningful way it is helpful to remind oneself
about some basic facts obtained by the evaluation of prime graphs:
graphs which are primitive under the coproduct and hence free of
subdivergences. They play the role of primes underlying the sought
after factorization and provide a rich source of number-theoretic
structure in quantum physics. Hence we briefly  review the role of
number theory in connection with residues in quantum field
theory. This is certainly one of the most surprising subjects
worthy of study in quantum field theory: the intimate connection
between transcendence and number theory, the topology of Feynman
graphs and gauge symmetries has slipped our attention far too
long, but slowly is becoming a prominent theme in physics and
mathematics \cite{Goncharov,Book}. We will review the main results
and briefly comment on common structures between generalized
polylogs and Feynman graphs. We then continue to discuss the
factorization of QFT.

The material in section one is a review following
\cite{dennis,review}, the material in section two is, at least
partially, new or a report on work in progress.

\section{Lie and Hopf algebras of Feynman graphs} Feynman graphs
form a pre-Lie algebra. This result needs no more than tracing
through the basic definitions used in perturbation theory. The
first ingredient is a definition of $n$-particle irreducible
graphs: an $n$-particle irreducible graph ($n$-PI graph) $\Gamma$
consists of edges and vertices  such that upon removal of any set
of $n$ of its edges it is still connected. Its set of edges is
denoted by $\Gamma^{[1]}$ and its set of vertices is denoted by
$\Gamma^{[0]}$. Edges and vertices can be of various different
types.

The pre-Lie product defined below maps 1PI graphs to 1PI graphs,
and is thus a well-defined operation on such graphs. For any
vertex $v$ $\in \Gamma^{[0]}$ we call the set
$f_v:=\{f\in\Gamma^{[1]}\mid v\cap f\not=\emptyset \} $ its type.
It is  a set of edges. Edges of a graph are either internal or
external. If we shrink all internal edges to a point, we call the
resulting edge or vertex graph a residue: we define ${\bf
res}(\Gamma)$ to be the result of identifying
$\Gamma^{[0]}\cup\Gamma^{[1]}_{\rm int}$ with a point in $\Gamma$.
Under the Feynman rules, ${\bf res}(\Gamma)$ evaluates to the
corresponding tree-level contribution.

A pre-Lie product on graphs emerges when we start plugging graphs
into each other: an internal edge or a vertex is replaced by a 1PI
graph which has external edges which match the vertex or internal
edge to be replaced. Note that this incorporates a statement about
renormalizability: the graphs to be inserted should have a residue
which appears as a local interaction vertex. For a renormalizable
field theory, the superficially divergent graphs certainly fulfil
this criterion.

\subsection{The Pre-Lie Structure}
Consider two graphs $\Gamma_1,\Gamma_2$. First, assume that
$\Gamma_2$ is an interaction graph so that it has more than two
external legs. For a chosen vertex $v_i\in\Gamma_1^{[0]}$ such
that $f_{v_i}\sim \Gamma^{[1]}_{2,\rm ext}$ (indicating that the
two (multi-)sets are identical), we define
\be\Gamma_1\ast_{v_i}\Gamma_2=\Gamma_1/v_i\cup\Gamma_2/\Gamma^{[1]}_{2,\rm
ext},\ee where in the union of these two sets we create a new
graph by gluing each edge $f_j\in f_{v_i}$ to one element in
$\Gamma_{2,\rm ext}^{[1]}$. Then we sum over all these possible
bijections between $f_{vi}$ and $\Gamma_{2,\rm ext}^{1}$, and
normalize such that topologically different graphs are generated
precisely once.

We now define
\be\Gamma_1\ast\Gamma_2=\sum_{w\in\Gamma_1^{[0]}\atop f_w\sim
\Gamma^{[1]}_{2,\rm ext}} \Gamma_1\ast_w\Gamma_2.\ee All this can
be easily generalized to the insertion of self-energy graphs,
graphs which have just two external edges, replacing internal
edges by self-energy graphs which have the corresponding external
edges \cite{RHI,InsElim}. One then has:
\begin{theorem} {\rm \cite{RHI,DK1,InsElim}}
The operation $\ast$ is pre-Lie: \be [\Gamma_1\ast \Gamma_2]\ast
\Gamma_3  -  \Gamma_1\ast[\Gamma_2\ast \Gamma_3] = [\Gamma_1\ast
\Gamma_3]\ast \Gamma_2  - \Gamma_1\ast[\Gamma_3\ast \Gamma_2].\ee
\end{theorem}
To understand this theorem, note that the equation says that the
lack of associativity in the bilinear operation $\ast$ is
invariant under permutation of the elements indexed $2,3$. This
suffices to show that the antisymmetrization of this map fulfils a
Jacobi identity. Hence we get a Lie algebra ${\cal L}$ by
antisymmetrizing this operation: \be
[\Gamma_1,\Gamma_2]=\Gamma_1\ast\Gamma_2-\Gamma_2\ast\Gamma_1,\label{Lie}\ee
and a Hopf algebra ${\cal H}$ as the dual of the universal
enveloping algebra of this Lie algebra, on general grounds
\cite{RHI,CK1}. Typically, one restricts attention to graphs which
are superficially divergent, with residues corresponding to field
monomials in the Lagrangian, while superficially convergent graphs
can be incorporated using suitable semi-direct products with
abelian algebras \cite{RHI}. Fig.(\ref{f2}) gives examples of Lie
brackets for various different theories. Our notation here is
somewhat loose; an appropriate orientation of fermion lines in the
QED case is to be understood in the figure. Also, the sum over all
bijections ensures the correct summation over all orientations in
internal fermion loops. \bookfig{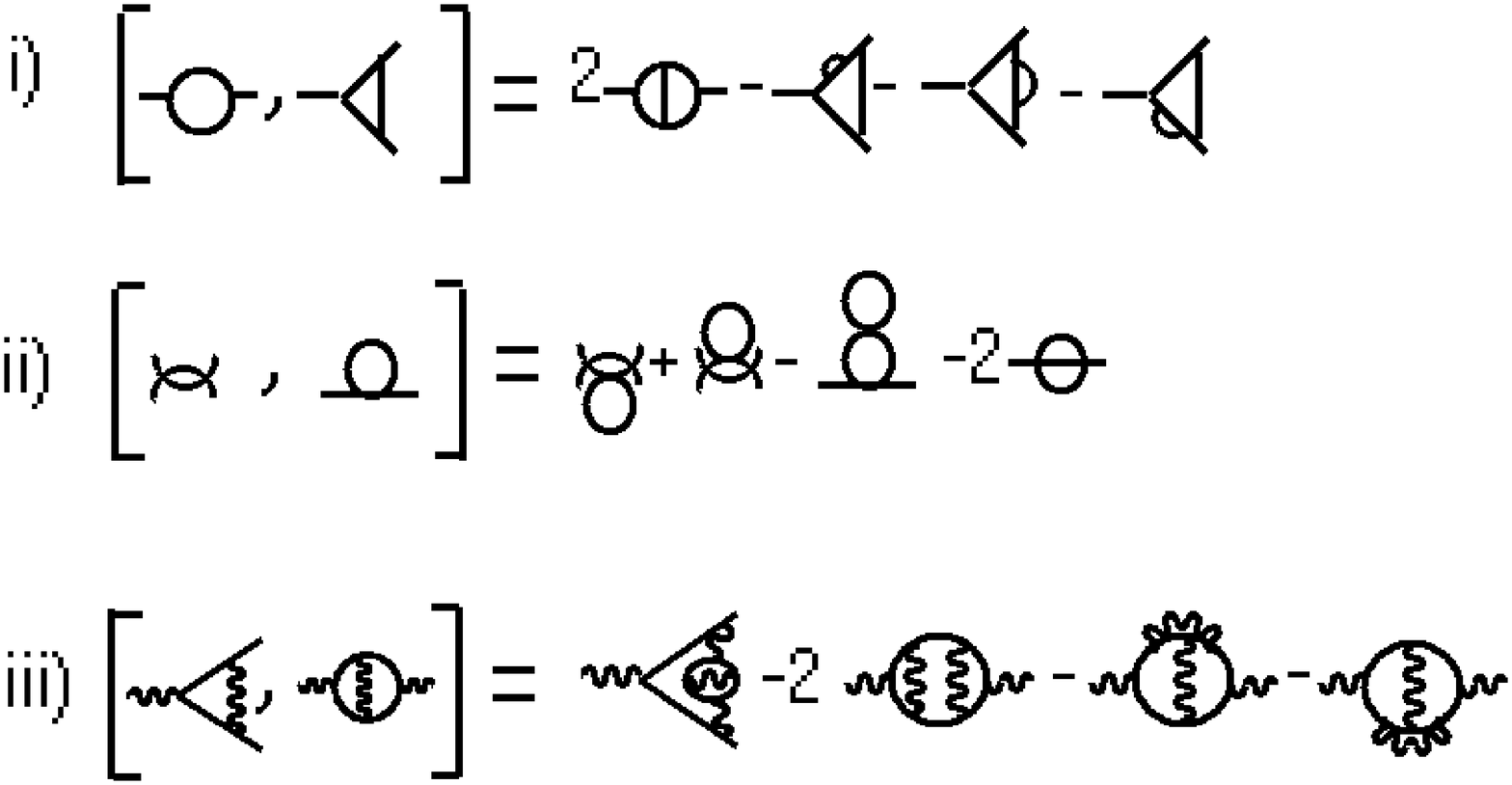}{f2}{f2}{Assorted Lie
brackets as examples: i) $\phi^3_6$ graphs, ii) $\phi^4_4$ graphs,
iii) QED graphs.}{6}

Similarly, if form-factor decompositions are needed this can be
incorporated using the notion of external structures or simply
colorings of (sub-) graphs \cite{RHI,dennis}.

\subsection{The principle of multiplicative subtraction}
Having defined Lie algebra structures on graphs, it is now easy to
harvest them to derive the renormalization process. As announced,
we just have to dualize the universal enveloping algebra ${\cal
U(L)}$ of ${\cal L}$ and obtain a commutative, but not
cocommutative Hopf algebra ${\cal H}$ \cite{RHI}. To find this
dual, one uses a Kronecker pairing and constructs it in accordance
with the Milnor-Moore theorem \cite{RHI,InsElim,CK1}.

We want to distinguish carefully now between the Hopf and Lie
algebras of Feynman graphs, so we write $\delta_\Gamma$ for the
multiplicative generators of the Hopf algebra and write $Z_\Gamma$
for the dual basis of the universal enveloping algebra of the Lie
algebra ${\cal L}$ with pairing \be \langle
Z_\Gamma,\delta_{\Gamma^\prime}\rangle=\delta^K_{\Gamma,\Gamma^\prime},\ee
where on the rhs we have the Kronecker $\delta^K$, and extend the
pairing by means of the coproduct \be\langle
Z_{\Gamma_1}Z_{\Gamma_2},X\rangle =\langle Z_{\Gamma_1}\otimes
Z_{\Gamma_2},\Delta(X) \rangle.\ee

First of all, we start by considering one-particle irreducible graphs
as the linear generators of the Hopf algebra, with their disjoint
union as product. We then identify  the  Hopf algebra as described
above  by a coproduct $\Delta:{\cal H}\to {\cal H}\otimes{\cal
H}$: \be\Delta(\Gamma)=\Gamma\otimes
1+1\otimes\Gamma+\sum_{\gamma{\subset}
\Gamma}\gamma\otimes\Gamma/\gamma,\ee where the sum is over all
unions of one-particle irreducible (1PI) superficially divergent
proper subgraphs and we extend this definition to products of
graphs so that we get a bialgebra. The above sum should, when
needed, also run over appropriate projections to formfactors, to
specify the appropriate type of local insertion \cite{RHI} which
appear in local counterterms, which we omitted in the above sum
for simplicity. Fig.(\ref{f3}) gives  examples of coproducts for
various theories. \bookfig{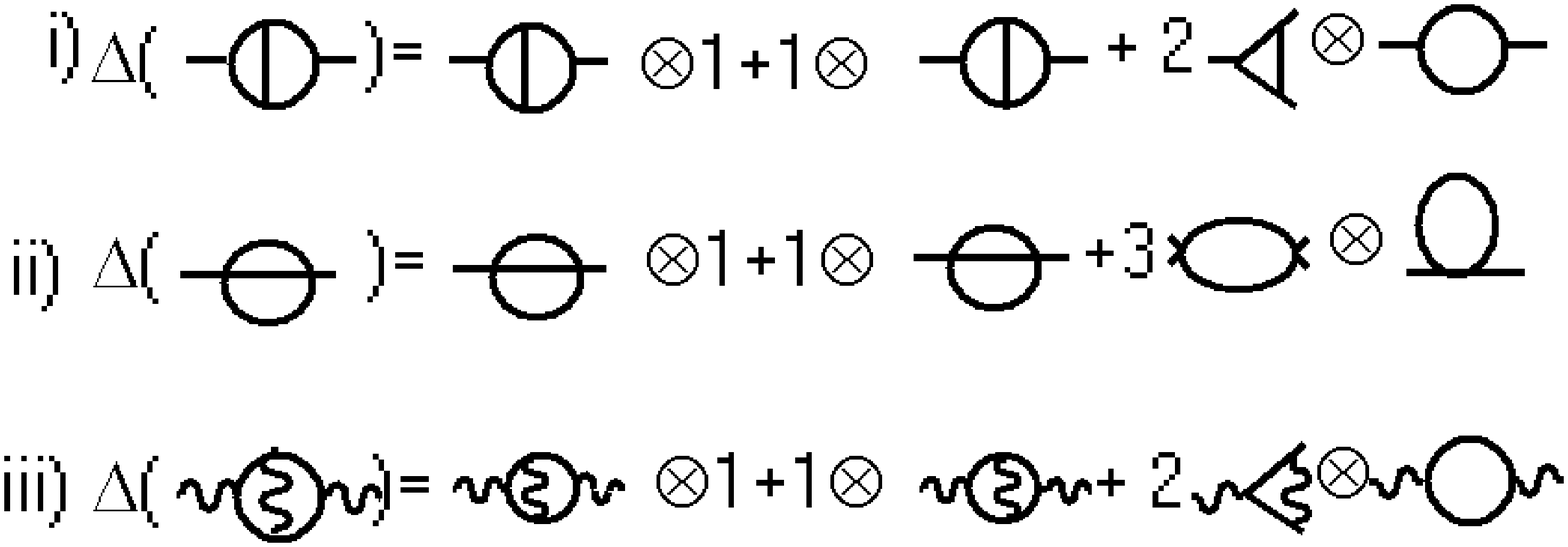}{f3}{f3}{Assorted coproducts
$\Delta(\Gamma)$: i) $\phi^3_6$, ii) $\phi^4_4$, iii) QED.}{4.7}

A short remark on notation: for any Hopf algebra element $X$ we
often write a shorthand for its coproduct $$
\Delta(X)=\widetilde{\Delta}(X)+X\otimes 1+1\otimes X=X\otimes
1+1\otimes X+X^\prime\otimes X^{\prime\prime}.$$ Let now $X$ be a
1PI graph. For each term in the sum $\widetilde{\Delta}(X)=\sum_i
X_{(i)}^\prime\otimes X_{(i)}^{\prime\prime}$ we have unique
gluing data $G_i$ such that \be
X=X_{(i)}^{\prime\prime}\ast_{G_i}X_{(i)}^\prime,\;\forall
i.\label{glue}\ee These gluing date describe the necessary
bijections to glue the components $X_{(i)}^\prime$ back into
$X_{(i)}^{\prime\prime}$ so as to obtain $X$: given the right
gluing data, we can reassemble the whole from its parts.

Having a coproduct, we still need a counit and antipode
(coinverse): the counit $\bar{e}$ vanishes on any non-trivial Hopf
algebra element, $\bar{e}(1)=1,\,\bar{e}(X)=0$. At this stage we
have a commutative, but typically not cocommutative bialgebra
\cite{Kassel}. It actually is a Hopf algebra as the antipode in
such circumstances comes for free as \be
S(\Gamma)=-\Gamma-\sum_{\gamma{\subset}
\Gamma}S(\gamma)\Gamma/\gamma.\ee

The next thing we need are Feynman rules, maps $\phi:{\cal H}\to
V$ from the Hopf algebra of graphs ${\cal H}$ into an appropriate
space $V$.

Over the years, we have invented many calculational schemes in
perturbative quantum field theory, and hence it is of no surprise
that there are many choices for this space. In any case, we will
have for disjoint 1PI graphs
$\phi(\Gamma_1\Gamma_2)\equiv\phi(\Gamma_2\Gamma_1)=\phi(\Gamma_1)\phi(\Gamma_2)$,
$\forall \phi:{\cal H}\to V$, where $V$ is an appropriate target
space for the evaluation of the graphs. Then, with the Feynman
rules providing a canonical character $\phi$, we will have to make
one further choice: a renormalization scheme. This is is a map
$R:V\to V$, and we demand that is does not modify the UV-singular
structure, and furthermore should obey \be
R(xy)+R(x)R(y)=R(R(x)y)+R(xR(y)),\label{RB}\ee
 an equation which guarantees the
multiplicativity of renormalization and is at the heart of the
Birkhoff decomposition to be discussed below: it tells us that
elements in $V$ split into two parallel subalgebras given by the
image and kernel of $R$ \cite{Chen}.  Algebras for which such a map
exists are known as Rota--Baxter algebras, a subject of increasing
importance recently \cite{Loday, Kurusch}.

Finally, the principle of multiplicative subtraction emerges: we
define a further character $S_R^\phi$ which deforms $\phi\circ S$
slightly and delivers the counterterm for $\Gamma$ in the
renormalization scheme $R$: \be
S_R^\phi(\Gamma)=-R[\phi(\Gamma)]-R\left[\sum_{\gamma{\subset}
\Gamma}S_R^\phi(\gamma)\phi(\Gamma/\gamma)\right]\ee which should
be compared with the undeformed \be \phi\circ
S=-\phi(\Gamma)-\sum_{\gamma{\subset} \Gamma}\phi\circ S
(\gamma)\phi(\Gamma/\gamma).\ee Then, the classical results of
renormalization theory follow immediately \cite{DK1,CK1,overl}. We
obtain the renormalization of $\Gamma$ by the application of a
renormalized character $$\Gamma\to S_R^\phi\star\phi(\Gamma)$$ and
Bogoliubov's $\bar{R}$ operation as
\be\bar{R}(\Gamma)=\phi(\Gamma)+ \sum_{\gamma{\subset}\Gamma}
S_R^\phi(\gamma)\phi(\Gamma/\gamma),\ee so that we have \be
S_R^\phi\star\phi(\Gamma)=\bar{R}(\Gamma)+S_R^\phi(\Gamma).\ee
Here, $S_R^\phi\star\phi$ is an element in the group of characters
of the Hopf algebra, with the group law given by
$$\phi_1\star\phi_2=m_V\circ(\phi_1\otimes\phi_2)\circ\Delta,$$
so that the coproduct, counit and coinverse (the antipode) give
the product, unit and inverse of this group, as  befits a Hopf
algebra. This Lie group has the previous Lie algebra ${\cal L}$ of
graph insertions as its Lie algebra \cite{RHI}.

In the above, we have given all formulas in their recursive form.
Zimmermann's original forest formula solving this recursion is
obtained when we trace our considerations back to the fact that
the coproduct of rooted trees can be written in non-recursive
form, and similarly the antipode \cite{overl}.  We also note that
the principle of multiplicative subtraction can be formulated in
much greater generality, as it is a basic combinatorial principle;
see for example \cite{Fot} for another appearance of this
principle.

\subsection{The Bidegree} A fundamental notion is the bidegree of
a 1PI graph. Usually, induction in perturbative QFT, aiming to
prove a desired result, is carried out using induction over the
loop number, an obvious grading for 1PI graphs. On quite general
grounds, for our Hopf algebras there exists another grading, which
is actually much more useful. We call it the bidegree, ${\rm
bid}(\Gamma)$ \cite{dennis,BK}. To motivate it, consider a
superficially divergent $n$-loop graph $\Gamma$ which has no
divergent subgraph. It is evident that its short-distance
singularities can be treated by a single subtraction, for any $n$.
It is not the loop number, but the number of divergent subgraphs,
which is the most crucial notion here. Fortunately, this notion
has a precise meaning in the Hopf algebra of superficially
divergent graphs using the projection into the augmentation ideal,
a projection which has the scalars $q1\equiv qe$ as its kernel. This indeed
counts the degree in renormalization parts of a graph: an overall
superficially convergent graph has bidegree zero by definition, a
primitive Hopf algebra element has bidegree one, and so on.

So we have ${\cal H}=\bigoplus_{i=0}^\infty{\cal H}^{(i)}$, with
${\rm bid}({\cal H}^{(i)})=i$.   To define this decomposition, let
${\cal H}_{\rm Aug}$ be the augmentation ideal of the Hopf
algebra, and let $P: {\cal H}\to {\cal H}_{\rm Aug}$ be the
corresponding projection $P={\rm id}-E\circ \bar{e}$, with
$E(q)=q1\in {\cal H}$. Let
$\widetilde{\Delta}(X)=\Delta(X)-1\otimes X-X\otimes 1$, as
before. $\widetilde{\Delta}$ is still coassociative, and for any
$X\in {\cal H}_{\rm Aug}$ there exists a unique maximal $k$ such
that $ \widetilde{\Delta}^{k-1}(X)\in [{\cal H}^{(1)}]^{\otimes
k}.$ Here, ${\cal H}^{(1)}$ is the linear span of primitive
elements $y$: $\Delta(y)=y\otimes 1+1\otimes y$. We call this
maximal $k$ the bidegree of a graph $\Gamma$.

As an example, the reader might determine the bidegree of the
graphs in Figs.(\ref{f2},\ref{f3}) and can check that it is
homogeneous under the Lie bracket as well as under the coproduct
and under the product (disjoint union). Typically, all properties
connected to questions of renormalization theory can be proven
more efficiently using the grading by the bidegree instead of the
loop number, a point which deserves some detailed comment.
\subsection{Renormalization and Hochschild Cohomology}
Each Feynman graph $\Gamma$ can be written in the form
$\Gamma=B_+^{\gamma,G_X}(X)$, where $\gamma$ is a bidegree one
graph, $X$ is a collection of subdivergences of $\Gamma$ such
that, when we shrink them all to a point in $\Gamma$, $\gamma$
remains, and $G_X$ is some data which tells us where to insert
these subdivergences. Any such map $B_+^{\gamma,G_X}$ extends to a
map on the Hopf algebra which is a closed Hochschild one-cocycle
\cite{dennis,CK1}.

This suggests a particularly nice way to prove locality of
counterterms and finiteness of renormalized Green functions,  by
using the Hochschild closedness of the operator
$B_+^{\gamma,G_X}$. Indeed it raises the bidegree by one unit and
is therefore a natural candidate to obtain such bounty. Underlying
this approach is the kinship between the Hopf algebras of Feynman
graphs with the universal Hopf algebra of non-planar rooted trees,
which has a very simple Hochschild cohomology \cite{CK1,Foissy}.

We will proceed by an induction over the bidegree, which is much
more natural than the usual induction over the number of loops. So
assume that $S_R\star\phi(\Gamma)$ is finite  and $S_R(\Gamma)$ a
local counterterm for all $\Gamma$ with ${\rm bid}(\Gamma)\leq k$.
Show these properties for all $\Gamma$ with ${\rm
bid}(\Gamma)=k+1$.

The start of the induction is easy: at unit bidegree,
$\phi(\Gamma)-R[\phi(\Gamma)]$ is finite and $S_R(\Gamma)$ is
local by assumption on $R$.

Let us assume we have established the desired properties of $S_R$
and $S_R\star\phi$ acting on all Hopf algebra elements up to
bidegree $k$. Assume ${\rm bid}(\Gamma)=k+1$. We have \be
\Gamma=B_+^{\gamma,G}(X),\ee where ${\rm bid}(\gamma)=1$, ${\rm
bid}(X)=k$, $X$ some monomial in the Hopf algebra.

Next, \be\Delta(\Gamma)=B_+^{\gamma,G}(X)\otimes 1+\left[1\otimes
B_+^{\gamma,G}\right]\Delta(X),\ee which expresses the crucial
fact that $B_+^{\gamma,G}$ is a closed Hochschild one-cocycle.

Using the Hochschild closedness of $B_+^{\gamma,G}$ one
immediately gets \be S_R\star\phi(\Gamma)=S_R(\Gamma)+{\bf
B}_+(\phi;S_R\star\phi;\gamma,G;X)\ee and \be S_R(\Gamma)= -R[{\bf
B}_+(\phi;S_R\star\phi;\gamma,G;X)].\ee Here we use a map ${\bf
B}_+(\phi;S_R\star\phi;\gamma,G;X)$ which inserts the renormalized
results $S_R\star\phi$ into the integral $\phi(\gamma)$ in
accordance with the gluing data \cite{Chen,dennis}.

From here, the induction step boils down to a simple estimate
using the fact that the powercounting for asymptotically large
internal loop momenta in $\phi(\gamma)$ is modified by the
insertion of $S_R\star\phi(X)$ (which is finite by assumption,
having bidegree $k$) only by powers of logarithms of internal
momenta of $\gamma$, and that delivers the result easily, using
the standard integral representation by the Feynman rules \be
\phi(\Gamma)=\int \prod_{e\in \Gamma^{[1]}_{\rm int}} d^D
k_{e}\;P^{-1}(k_{e})\prod_{v\in\Gamma^{[0]}}\delta^{(D)}\left(\sum_{j\in
f_v}k_j\right) g(v),\label{int}\ee with a suitable ordering of
propagators and vertices understood. A finite renormalization to
achieve not only finiteness, but for example to resurrect the
gauge invariance of the theory, can be incorporated in this
approach via a further convolution with a character of the Hopf
algebra. Details of such an approach will be the subject of future
work.

This ends the review of the basic notions of renormalization
theory. It remains to comment on progress which was initiated by
this algebraic viewpoint along two lines: a connection to the
Riemann--Hilbert problem \cite{RHI,RHII} and strong hints towards
connections with number theory, coming from the values of residues
of bidegree one graphs \cite{Book}, as well as from the structure
of the Dyson--Schwinger equations, but also arising from number theory
itself \cite{Connes}. But first, let us review the connection to
the Riemann--Hilbert problem.
\subsection{The Birkhoff decomposition and the renormalization
group} Where do we stand now? We have recognized the iterative
subtraction mechanism of perturbative quantum field theory as a
Hopf algebra structure. The Bogoliubov recursion designed to
guarantee local counterterms originates in very natural Lie and
Hopf algebra structures of graphs, and thus forest formulas have
been given their mathematical identification. The Lie group of
characters on this Hopf algebra is based on a rather huge Lie
algebra of antisymmetrized graph insertions. It has as many
generators as there are 1PI graphs, and even if we restrict
ourselves to the primitive (bidegree one) graphs into which any
graph decomposes, we still are confronted with an infinite number
of those, if our theory is renormalizable. Still, the algebraic
structures reported so far allow for surprising new insight into
the structure of QFT. A first such step is the recognition of the
algebraic constraint on the renormalization map $R$. It leads to a
Birkhoff decomposition which relates QFT to the Riemann--Hilbert
problem \cite{RHI,RHII}. This certainly gives hope for a better understanding of
the analytic structure of Green functions, as they now
start looking like a generalization of other solutions to a
Riemann--Hilbert problem, with KZ equations and hypergeometric
functions coming to mind.

Further progress was made upon recognition of the role
diffeomorphisms of physical parameters play in this context: group
homomorphisms from the group of characters of Feynman graphs to
diffeomorphisms of physical parameters are provided galore by QFT,
and the Birkhoff decomposition is compatible with these
homomorphisms: an unrenormalized physical observable has a
decomposition into a bare and a renormalized part, a result which
summarizes in one line the wisdom of locality and the
renormalization group \cite{RHII}. Still, the link towards the
Riemann--Hilbert problem reveals the deficiencies of perturbative
quantum field theory quite pointedly: the decomposition makes
sense only in an infinitesimal disk, the order of the pole is
unbounded and the diffeomorphism is anyhow only a formal one. The
latter point cries for resummation, and the former points, as we will
argue, demand some renormalization group improvement of
perturbation theory, based on a factorization of graphs to be
discussed below, to restore the credibility of perturbation theory
as an input in any means to come to conclusions on the
non-perturbative theory.

The Feynman rules in dimensional or analytic regularization
determine a character $\phi$ on the Hopf algebra which evaluates
as a Laurent series in a complex regularization parameter $\ve$,
with poles of finite order, this order being bounded by and hence
dependent on the bidegree of the Hopf algebra element to which
$\phi$ is applied. In minimal subtraction, $\phi_-:=S^\phi_{R=MS}$
has similar properties: it is a character on the Hopf algebra
which evaluates as a Laurent series in a complex regularization
parameter $\ve$, with poles of finite order, this order being
bounded by the bidegree of the Hopf algebra element to which
$S^\phi_{R=MS}$ is applied, only that there will be no powers of
$\ve$ which are $\geq 0$. Then, $\phi_+:=S^\phi_{R=MS}\star\phi$
is a character which evaluates in a Taylor series in $\ve$; all
poles are eliminated. We have the Birkhoff decomposition
\be\phi=\phi_-^{-1}\star\phi_+.\ee

This establishes  an amazing connection between the
Riemann--Hilbert problem and renormalization \cite{RHI,RHII}. It
uses in a crucial manner once more that the multiplicativity
constraints Eq.(\ref{RB}),
$$ R[xy]+R[x]R[y]=R[R[x]y]+R[xR[y]],$$ ensure that the
corresponding counterterm map $S_R$ is a character as well, \be
S_R[xy]=S_R[x]S_R[y],\;\forall x,y\in H,\ee by making the target
space of the Feynman rules into a Rota--Baxter algebra,
characterized by this multiplicativity constraint. The connection
between Rota--Baxter algebras and the Riemann--Hilbert problem,
which lurks in the background here, remains largely unexplored as
of today.

As announced, renormalization in the MS scheme can now be
summarized in a single phrase: with the character $\phi$ given by
the Feynman rules in a suitable regularization scheme and
well-defined on any small curve around $\ve=0$, find the Birkhoff
decomposition $\phi_+(\ve)=\phi_-\star\phi$.

 The unrenormalized analytic
expression for a graph $\Gamma$ is then $\phi[\Gamma](\ve)$, the
MS-counterterm is $S_{{ MS}}(\Gamma)\equiv\phi_-[\Gamma](\ve)$ and
the renormalized expression is the evaluation $\phi_+[\Gamma](0)$.
Once more, note that the whole Hopf algebra structure of Feynman
graphs is present in this group: the group law demands the
application of the coproduct, $\phi_+=\phi_-\star \phi\equiv
S_{MS}^\phi\star\phi$.

But still, one might wonder what a huge group this group of
characters really is. What one confronts in QFT is the group of
diffeomorphisms of physical parameters: lo and behold, changes of
scales and renormalization schemes are just such (formal)
diffeomorphisms. So, for the case of a massless theory with one
coupling constant $g$, for example, this just boils down to formal
diffeomorphisms of the form $$g\to \psi(g)=g+c_2 g^2+\ldots.$$ The
group of one-dimensional diffeomorphisms of this form looks much
more manageable than the group of characters of the Hopf algebras
of Feynman graphs of such a theory.

\subsection{Diffeomorphisms of physical parameters}
Thus, it would be very nice if the whole Birkhoff decomposition
could be obtained at the level of diffeomorphisms of the coupling
constants. This is certainly most desirable from a physicists'
viewpoint: after all, we would like to have the theory
parametrized by physical observables, and changes we can make in
our way of formulating the theory should correspond to changes we
can make in those observables.

The crucial step toward that goal is to realize the role of a
standard QFT formula of the form (in the context of $\phi^3_6$
theory, say) \be g_{\rm new}=g_{\rm old}\;Z_1 Z_2^{-3/2},\ee which
expresses how to obtain the new coupling in terms of a
diffeomorphism of the old. This was achieved in \cite{RHII},
recognizing this formula as a Hopf algebra homomorphism from the
Hopf algebra of diffeomorphisms to the Hopf algebra of Feynman
graphs, regarding $Z_g=Z_1/\;Z_2^{3/2}$, a series over
counterterms for all 1PI graphs with the external leg structure
corresponding to the coupling $g$, in two different ways. It is at
the same time a formal diffeomorphism in the coupling constant
$g_{\rm old}$ and a formal series in Feynman graphs. As a
consequence, there are two competing coproducts acting on $Z_g$.
That both give the same result defines the required homomorphism,
which transposes to a homomorphism from the largely unknown group
of characters of ${\cal H}$ to the one-dimensional diffeomorphisms
of this coupling.

The crucial fact in this is the recognition of the Hopf algebra
structure of diffeomorphisms by Connes and Moscovici \cite{CM}:
Assume you have formal diffeomorphisms $\phi,\psi$ in a single
variable \be x\to\phi(x)=x+\sum_{k>1}c_k^\phi x^k,\label{expa}\ee
and similarly for $\psi$. How do you compute the Taylor
coefficients $c^{\phi\circ\psi}_k$ for the composition
$\phi\circ\psi$ from the knowledge of the Taylor coefficients
$c_k^\phi,c_k^\psi$? It turns out that it is best to consider the
Taylor coefficients
\be\delta_k^\phi=\log(\phi^\prime(x))^{(k)}(0)\label{CMT}\ee
instead, which are as good to recover $\phi$ as the usual Taylor
coefficients. The answer lies then in a Hopf algebra structure:
\be\delta^{\phi\circ\psi}_k=m\circ(\tilde{
\psi}\otimes\tilde{\phi} )\circ\Delta_{CM}(\delta_k), \ee where
$\tilde{\phi},\tilde{\psi}$ are characters on a certain Hopf
algebra ${\cal H}_{CM}$ (with coproduct $\Delta_{CM}$) so that
$\tilde{\phi}( \delta_i)$ $=$ $\delta_i^\phi$, and similarly for
$\tilde{\psi}$. Thus one finds a Hopf algebra with abstract
generators $\delta_n$ such that it introduces a convolution
product on characters evaluating to the Taylor coefficients
$\delta_n^\phi,\delta_n^\psi$, such that the natural group
structure of these characters agrees with the diffeomorphism
group. This is a very small piece of the work in \cite{CM}, which
was very crucial though in understanding the connection between
the group of diffeomorphisms of physical parameters and the group
of characters on our Hopf algebra ${\cal H}$: it turns out that
this Hopf algebra of Connes and Moscovici is intimately related to
rooted trees in its own right \cite{CK1}, signalled by the fact
that it is linear in generators on the rhs, as are the coproducts
of rooted trees and graphs \cite{InsElim,CK1}.

 There
are a couple of basic facts which enable one to make in general
the transition from this rather foreign territory of the abstract
group of characters of a Hopf algebra of Feynman graphs (which, by
the way, equals the Lie group assigned to the Lie algebra with
universal enveloping algebra the dual of this Hopf algebra) to the
rather concrete group of diffeomorphisms of physical observables.
These steps are: \begin{itemize}
\item Recognize that $Z$ factors are given as counterterms over
formal series of graphs starting with 1, graded by powers of the
coupling, hence invertible.
\item Recognize  the series $Z_g$ as a formal diffeomorphism,
with Hopf algebra coefficients.
\item Establish that the two competing Hopf algebra structures of
diffeomorphisms and graphs are consistent in the sense of a Hopf
algebra homomorphism.
\item Show that this homomorphism transposes to a Lie algebra and
hence Lie group homomorphism.
\end{itemize}
This works out extremely well, with details given in \cite{RHII}.
In particular, the effective coupling $g_{\rm eff}(\ve)$ now
allows for a Birkhoff decomposition in the space of formal
diffeomorphisms: \begin{theorem}  {\rm \cite{RHII}} \be g_{\rm
eff}(\ve)=g_{\rm eff -}(\ve)^{-1}\circ g_{\rm eff +}(\ve) \ee
where $g_{\rm eff -}(\ve)$ is the bare coupling and $g_{\rm eff
+}(0)$ the renormalized effective coupling.\end{theorem} The above
results hold as they stand for any massless theory which provides
a single coupling constant, with the relevant Hopf algebra
homomorphism for example in the QED case given by $e_{\rm
new}=Z_3^{-1/2}e_{\rm old}$ (and $Z_3$ regarded as a sum over all
1PI vacuum polarization  diagrams). If there are multiple
interaction terms in the Lagrangian, one finds similar results
relating the group of characters of the corresponding Hopf algebra
to the group of formal diffeomorphisms in the multidimensional
space of coupling constants.

Finally, the Birkhoff decomposition of a loop, $ \d (\ve) \in {\rm
Diff} \, (X),$ admits a beautiful geometric interpretation
\cite{RHII}, described in Fig.(\ref{pic}).
\bookfig{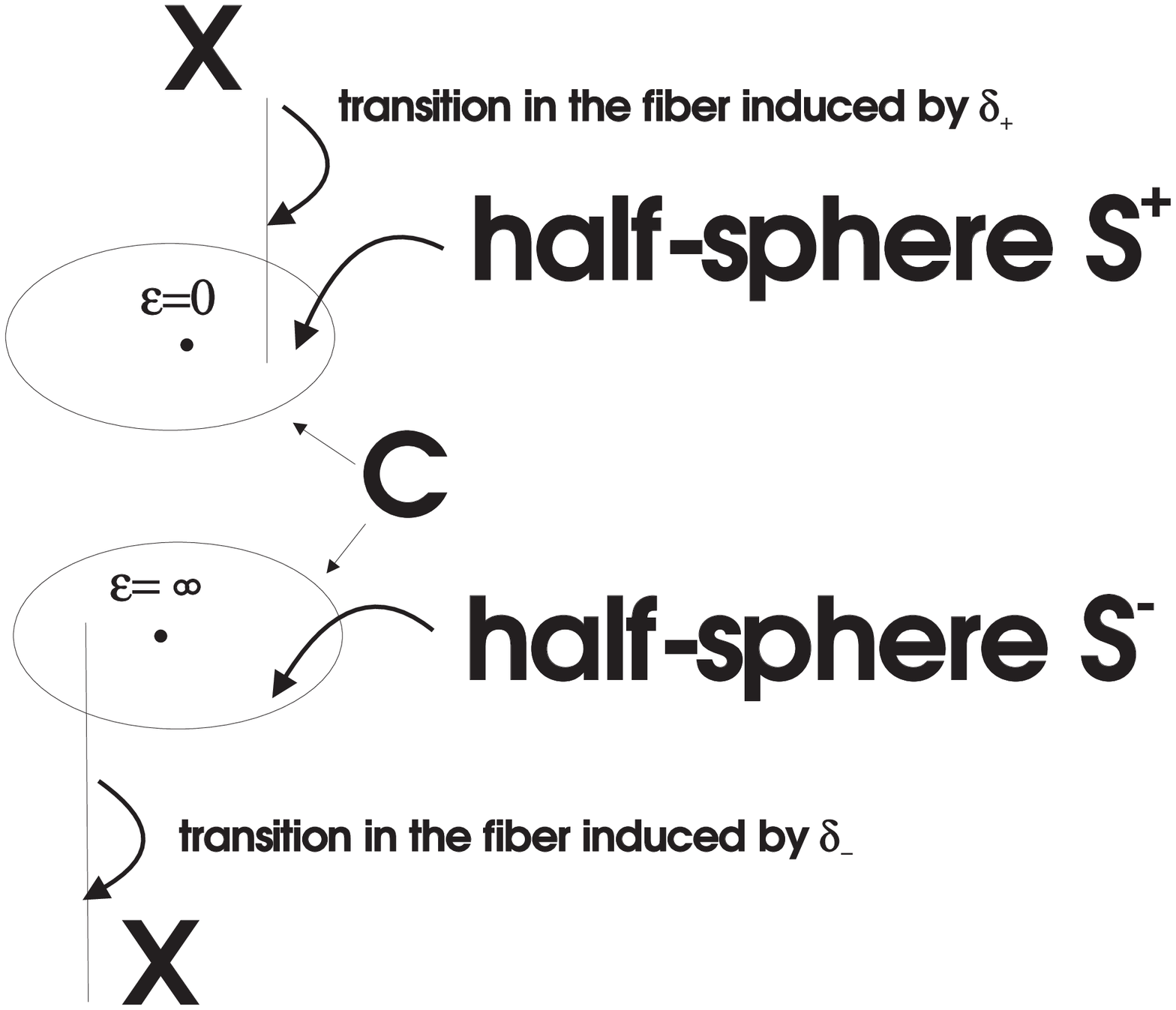}{}{pic}{A geometric picture for the Birkhoff
decomposition \cite{RHII}. Here, $\delta$ is the character
obtained from $\phi$ by evaluating it as a complex number on an
infinitesimal loop around the point of interest $\epsilon=0$, and
$\delta_\pm$ are the components of its Birkhoff decomposition
which induce transitions (formal diffeomorphisms) in the fiber
$X$.}{4}

So what stops us from using this connection to the Riemann--Hilbert
problem and establishing quantum field theory as a solution to this
problem? There are two topics here: first of all, we are up to now
talking about formal series, and a resummation is certainly needed
to turn our formal diffeomorphisms into actual ones. Here, recent
progress by Ramis \cite{Ramis} with formal series in connection
with the Riemann--Hilbert problem, even in the case of zero radius
of convergence, hopefully proves very relevant.  It is in
particular encouraging to see the emergence of "ambiguity groups"
\cite{Connes,Ramis} appearing in this context: a proper
identification of the renormalization group in terms of a Galois
symmetry is one of the ideas which has slowly emerged in recent years.

But even before resummation, for each term in the perturbation
series, the finite value is not necessarily the right input
parameter for such a resummation. There are well-known
deficiencies of perturbation theory \cite{Riv,Wight}:
\begin{itemize}
\item the subtraction of a counterterm in perturbation theory
renders ambiguous dependencies on logarithms of scales in the
renormalized amplitudes which are not to be trusted as such, and
in conflict with the requirements from the renormalization group.
A multiscale expansion seems to capture the essence of scaling in
QFT more faithfully. Nevertheless, the exactness of perturbation
theory is striking, and overcoming this obstacle without the
sacrifice of the achievements of momentum space Feynman diagram
perturbation theory would be most desirable.
\item Iterating chains of one-loop graphs can produce renormalons
in perturbation theory. On occasion, they can be used to
parametrize the unknown regime of the non-perturbative, but are in
the end just a suspicious infinite sum of the previous obstacle.
\item $S_R^\phi(\Gamma)$, for ${\rm bid}(\Gamma)>1$, is a
Laurent series which has poles of higher order, though all
subdivergences have been eliminated in that local counterterm. It
would be more natural, and desirable for our Riemann--Hilbert
decomposition, if the pole term would be only of first order, say,
after absorbing the subdivergences: a uniform bound, independent
of the bidegree of $\Gamma$, on the order of the pole term would
make our Riemann--Hilbert problem much more regular, even if the
coefficients of that finite order pole still form a series in the
coupling with vanishing radius of convergence. The appearance of
higher order poles is again related to the first obstacle, as they
arise from an iteration of scaling degrees coming from
subdivergences calculated in perturbation theory. These poles are
indeed completely determined by the residues in the theory
\cite{RHII}, and can be obtained from the scattering-type formula
of \cite{RHII}, with combinatorial coefficients which turn out to
be generalized factorials \cite{Chen,KD}, by that formula. These
poles are thus highly redundant and again reflect our inefficient
handling of scaling properties in perturbation theory.
\item at higher loop orders, poles appear which are arbitrarily
close to the region of interest (a little disk around
$\epsilon=0$), which typically come from the expansion of
$\Gamma(1\pm n\epsilon)$ in perturbation theory, with $n$ being
the loop number. Again, the appearance of these poles at
$\pm\;1/n$ can be traced back to the same origin as the previous
obstacles. These poles force us (for large loop number) to
consider an infinitesimal disk around $\epsilon=0$ in the Birkhoff
decomposition.
\end{itemize}
Alas, the logarithmic scaling properties of perturbation theory
are not in accordance with the exact renormalization group and to
overcome this difficulty, and to understand better the relation
between perturbative and non-perturbative approaches, again the
Lie algebra of Feynman graphs offers assistance. This is a very
new development, and we will in the next section just outline some
recent work in progress, partially mentioned already in
\cite{dennis}. We start by motivating factorizations in quantum
field theory.

\section{Perspective: Euler products in QFT}
In this section we want to comment on a connection between
Dyson--Schwinger equations and Euler products. Ultimately, I
believe that there is a deep connection between the two subjects,
and to motivate this connection let us start with a subject from
number theory, the Riemann $\zeta$ function, and obtain it as a
solution to a Dyson--Schwinger equation. For now, this is only
meant as a sufficient stimulus to invert the reasoning and look
for Euler products in quantum field theory.
\subsection{The Riemann $\zeta$ function from a Dyson--Schwinger
eq\-uation} The Riemann $\zeta$-function is the analytic
continuation of the sum $\sum_n 1/n^s$, and can be written in the
form of an Euler product \be
\zeta(s)=\sum_{n}\frac{1}{n^s}=\prod_p\frac{1}{1-p^{-s}},\; \Re
(s)
>1,
\ee where the product is over all primes $p$ of the (rational)
integers.

Let us now define a Hopf algebra of sequences $(p_1,\ldots,p_k)$,
where the $p_i$ are primes, and introduce $B_+^p[J]$ as the
sequence which is obtained by adding a new prime $p$ as the first
element to the sequence $J$, for example
$B_+^3[(5,3,2)]=(3,5,3,2)$. The Hopf algebra structure emerges
when we require that $B_+^p$ is Hochschild closed for all $p$: \be
\Delta(B_+^p[J])=B_+^p[J]\otimes 1+[{\rm id}\otimes
B_+^p]\Delta[J],\ee with $\Delta(1)=1\otimes 1$ and where we identify
$1$ with the empty sequence. Define the value $w(J)$ to be the
product of the entries of $J$, and let the symmetry factor $S(J)$
be $k!$ if the sequence has length $l(J)=k$, which avoids overcounting below.  
Note that for a one
element sequence $(p)$, \be \Delta[(p)]=(p)\otimes 1+1\otimes
(p),\ee primitive elements have prime value, $w((p))=p$.

Consider the "Dyson--Schwinger equation" \be
\overline{\zeta}(\rho)=1+\rho\sum_p
B_+^p[\overline{\zeta}(\rho)],\ee so that we obtain a formal
series (in "the coupling" $\rho$) \be
\overline{\zeta}(\rho)=1+\rho\sum_p
(p)+\rho^2\sum_{p_1,p_2}(p_1,p_2)+\cdots . \ee Define "Feynman
rules" by $\phi_s(J)=\frac{1}{l(J)!}w(J)^{-s}$, and set \be
\zeta(s,\rho)=\phi_s[\overline{\zeta}(\rho)]. \ee Then, we recover
Riemann's $\zeta$ function as \be \zeta(s)=\lim_{\rho\to
1}\zeta(s,\rho). \ee

Note the general structure of the formal ``Dyson--Schwinger
equation" above: it determines an unknown $\zeta(\rho)$ in terms
of itself, as "1 plus a sum over the image of the unknown
$\zeta(\rho)$ under all closed Hochschild one cocycles $B_+^p$,
weighted by appropriate symmetry factors".

Next, we remind ourselves that $\zeta(s)$ has an Euler product. Is
there an Euler product for $\overline{\zeta}$?

The answer is yes, and the simplest way is to get it from the
well-known shuffle product on sequences. We introduce this
associative and commutative product via \be B_+^{p_1}(J_1)\sqcup
B_+^{p_2}(J_2)=B_+^{p_1}(J_1\sqcup
B_+^{p_2}(J_2))+B_+^{p_2}(B_+^{p_1}(J_1)\sqcup J_2). \ee Then, \be
\overline{\zeta}(\rho)=\Pi_p^\sqcup\frac{1}{1-\rho\; (p)},
\;{\mbox{where}}\; \frac{1}{1-\rho\;(p)}=1+\rho\;(p)+\rho^2\;
(p)\sqcup(p)+\cdots,\ee and where the shuffle product is used in
the Euler product throughout. We then have
\be\zeta(s)={\phi_s}_{\mid_{\rho=1}}\left( \Pi_p^\sqcup\frac{1}{1-\rho\; (p)}\right)
=\Pi_p \frac{1}{1-p^{-s}},\ee the evaluation of the product is the
product of the evaluations.

The reason we dared calling the above equation a Dyson--Schwinger
equation is a simple fact - the true Dyson--Schwinger equations of
QFT have a similar structure: they express an unknown Green
function as a sum over all possible insertions of itself in all
possible skeleton diagrams. This allows us to write the unknown Green
function as a sum over all possible images over all closed
Hochschild one-cocycles in the theory (the $B_+^\gamma$ obtained
by summing over all possible gluing data $G_i$ in the
$B_+^{\gamma,G_i}$ considered earlier), precisely provided by the
primitive (bidegree one) graphs $\gamma$, which play the role of
primes. Let us review quickly their fascinating properties first.
\subsection{Residues in QFT}
Consider a Feynman graph in some renormalizable quantum field
theory and assume the graph is free of superficially divergent
subgraphs. We can always restrict ourselves to logarithmically
divergent graphs by factorizing out suitable polynomials in masses
and external momenta. Then, such a logarithmic divergent quantity
has a residue which is independent of all these parameters. It is
a well-defined number and the only chance we have of changing this
number is to change the topology of the graph under consideration.
So that should be a rather interesting number, and indeed, nature
rewards us for posing a good question by revealing an intimate
connection between the topology of the graph and the
number-theoretic residues one obtains upon evaluating such a
graph. The residue here is the coefficient of the short-distance
singularity in such a graph, calculated as the coefficient of the
first order pole in dimensional regularization, or even as a
residue in the operator-theoretic sense. As our graph has bidegree
one, it provides a residue which is  a universal number
independent of the choice of a regularization. Topologically, the
simplest graphs are ladder graphs. Their residues are rational
numbers \cite{Book}.

Then, the next class of graphs are graphs which have a less
trivial topology, reflected by a non-trivial Gauss code,
$(1,2,1,2)$ being the first such topology given in
Fig.(\ref{gauss}), see \cite{Book}. By all computational
experience, graphs which have such a Gauss code deliver a residue
$\sim \zeta(3)$. From there, a whole universe unfolds, revealing
deep connections between the symmetries in a QFT, and its
transcendental richness \cite{Book}.
\bookfig{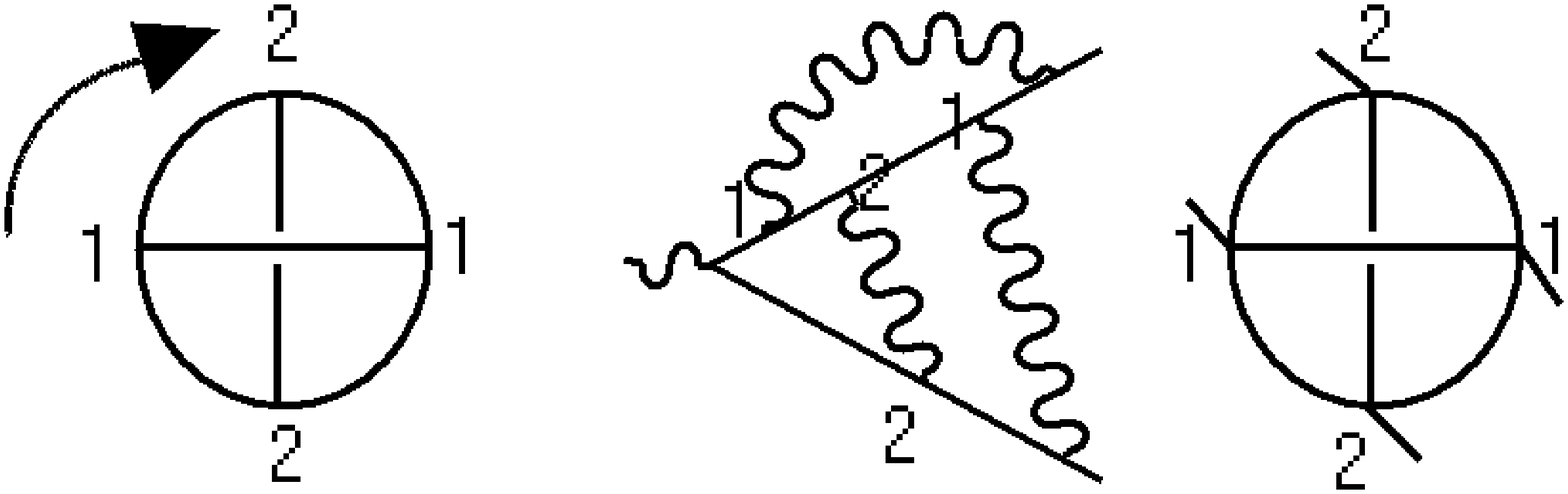}{}{gauss}{Non-exhaustive list of examples of
QFT graphs realizing the  Gauss code diagram $\{1,2,1,2\}$ (on the
left), related to the appearance of $\zeta(3)$ in their
evaluation. The two Feynman graphs are from $\phi^4_4$ and QED.
Only internal vertices matter.}{3}

One remarkable fact is that the decomposition into two-line
reducible parts corresponds to a factorization of graphs which is
compatible with their evaluation: the evaluation of the full graph
delivers the product of the evaluation of the parts, as in the
product of prime knots \cite{Book,BKold1,BKold2}.

There is no space here to comment on the weird and wonderful data
with which renormalizable QFT provide us in such circumstances,
with fascinating new phenomena appearing at higher loop orders
\cite{brosnan}, and we refer the reader to \cite{Book} for an
exhaustive census of such phenomena. But still, one fact is worth
mentioning: the relation between the presence or absence of
transcendental numbers depending on the internal symmetries in the
theory, a connection which started with Rosner's observation
\cite{Rosner} of the absence of $\zeta(3)$ in the residues of QED
at three loops, and which has found even more striking
confirmation ever since, but still deserves much further
exploration \cite{Book,BDK}.

Also, there are two basic structures in Feynman graphs: the
convolution of renormalization schemes \be
S_{R_1}^\phi\star\phi=[S_{R_1}^\phi\star S_{R_2}^\phi\circ S]\star
[S_{R_2}^\phi\star\phi],\ee which generalizes Chen's lemma
\cite{Chen}, and the generalized shuffle identity \be
\phi(\Gamma_1\vee\Gamma_2)\sim\phi(\Gamma_1)\phi(\Gamma_2),\ee the
factorization to be introduced below. In structure, they are very
similar to the relations which appear amongst generalized
polylogarithms \cite{Goncharov} and Euler--Zagier sums, the number
class most obviously related to Feynman diagrams, even if they
might not yet exhaust them. For now, radiative correction
calculations have stimulated many a development in that area of
number theory. Number theory in return hopefully is able to
further our understanding of QFT, in particular with respect to an
identification of a QFT by its transcendental nature, eventually.

\subsection{Factorizing graphs}
Let us now ask the question wether a factorization into Euler products
can be found in quantum field theory? And then, if this can be
found on the combinatorial level, will the evaluation, by the
Feynman rules, equal the product of the evaluations, and, if not,
by how much will it deviate?

After all, a typical Dyson--Schwinger equation is of the form \be
X=1+\sum_\gamma B_+^\gamma(g^k  [\cup_k X]),\ee where the infinite
sum in the Hopf algebra is over primitive graphs $\gamma$,
$k=k(\gamma)$ is the degree of $\gamma$, and as the notation
indicates, the maps $B_+^\gamma$ are closed Hochschild
one-cocycles, and the sum is over all of those. $X$ is here to be
regarded as an infinite sum of graphs contributing to a chosen
Green function, and evaluation by the Feynman rules delivers the
usual Dyson--Schwinger equations given as an integral equation
over the kernels provided by the primitive graphs $\gamma$. Note
that, as insertion into a primitive graph commutes with the
coproduct in the desired way, we can directly read off the
renormalized Dyson--Schwinger equation as \be X_R=Z_X+\sum_\gamma
B_+^\gamma(g^k  [\cup_k X_R]),\ee where $Z_X$ is the negative part
in the Birkhoff decomposition with respect to a renormalization
scheme $R$. Here, $[\cup_k X]$ indicates a $k$-fold disjoint union
of $X$, regarded as the product in the Hopf algebra of graphs.

Actually, we typically have a coupled set of such equations with
several unknowns (Green functions) but we here simply discuss the
structure of such equations, suitable generalizations being
straightforward.

So the natural question to ask is: is there an Euler product for
the formal sum generated by such an equation? The answer is indeed
affirmative.

 The crucial step lies in the definition of the
product $\vee$ which generalizes the shuffle product $\sqcup$,
appropriate for totally ordered sequences, to the partial order
given by being a subgraph.

Let us briefly describe this product: let a sequence of primitive
graphs $J=(\gamma_1,\ldots,\gamma_k)$ be given. We say that a
graph $\Gamma$ is compatible with that sequence, $\Gamma\sim J$,
iff its bidegree equals the length $k$ of the sequence and   $$
\langle Z_{\gamma_k}\otimes\ldots\otimes
Z_{\gamma_1},\tilde{\Delta}^{k-1}(\Gamma)\rangle\not= 0,$$ where
we use the previous pairing between the Lie algebra elements
$Z_\gamma$ and the Hopf algebra. Let $n_\Gamma$ be the number of
sequences compatible with $\Gamma$. Define
\be\Gamma_1\vee\Gamma_2=\sum_{{I_1\sim \Gamma_1\atop
I_2\sim\Gamma_2}} \sum_{\Gamma\sim I_1\sqcup
I_2}\frac{1}{n_\Gamma}\Gamma, \ee where the first sum is over all
sequences compatible with the two graphs $\Gamma_1,\Gamma_2$ and
the second sum is over all sequences appearing in the shuffle of
$I_1,I_2$, and over all compatible graphs $\Gamma$. This is a
commutative associative product on 1PI graphs. It has a relation
to the pre-Lie product introduced earlier, to be described
elsewhere. Then, we have
\begin{theorem}$X=\prod^\vee_\gamma \frac{1}{1-g^{k(\gamma)}\gamma},$\end{theorem}
the proof of which is elementary given the definition of the
product $\vee$, which maps 1PI graphs to 1PI graphs.

Most urgently needed is an understanding to what extent this is compatible
with the evaluation by Feynman rules $\phi$: how much can we say
about
$$ \phi\left(\Pi^\vee_\gamma
\frac{1}{1-\gamma}\right)\; \mbox{vs}\; \Pi_\gamma
\frac{1}{1-\phi(\gamma)}=\zeta_G(\phi)\;?$$ Here, $\zeta_G(\phi)$
shall be regarded as a ``$\zeta$ function" (in quotes, as we do not
give here any non-trivial results  concerning functional relations
or such) which, for a fixed Green function $G$ has an Euler
product over the primitive (bidegree one) graphs $\gamma$ (which
all have a graphical residue ${\rm \bf res}(\gamma)$ which agrees
with the tree level contribution to $G$) and where the variable
$\phi$ is the chosen character on the Hopf algebra of graphs
underlying the QFT in which the Green function appears.

To phrase it otherwise, what stops us from actually considering an
Euler product over all primitive graphs to get a formal solution
to the Dyson--Schwinger equations in general? Can we just construct
$\zeta$-functions associated to a chosen Green function, defined
via an Euler product over primitive elements?

A few comments are immediate: no, perturbation theory does not
factorize straightforwardly into its primitives. But there are
many encouraging signs. First of all, the scattering type formulas
of \cite{RHII} show that in dimensional regularization the
leading coefficient of the singularity respects the desired
factorization. This is useful. Indeed, for arbitrary superficially
divergent graphs $\Gamma_1,\Gamma_2$ one immediately shows \be
\frac{\phi(\Gamma_1\ast_v\Gamma_2))}{\phi(\Gamma_1)\phi(\Gamma_2)}=\frac{n_1+n_2}{n_2}(1+{\cal
O}(\epsilon)),\ee where $n_1,n_2$ are the numbers of loops in
$\Gamma_1,\Gamma_2$ and $\epsilon$ is the dimensional
regularization parameter (similarly in other regularizations).

The combinatorial pre-factor $(n_1+n_2)/n_2$ is easy to understand
and to deal with. It is in the non-leading terms where progress
had to be made. But let us muse a bit about what the consequences
of such a factorization would be. Using the definition of
$S_R^\phi$, one immediately has, for products of primitives,
\bea\phi(\gamma_1\ast_v\gamma_2)=\phi(\gamma_1)\phi(\gamma_2) &
\Leftrightarrow & S_R^\phi(\gamma_1\ast_v\gamma_2)\nonumber\\  =
-R\left[
\phi(\gamma_1)\phi(\gamma_2)-R[\phi(\gamma_2)]\phi(\gamma_1)\right]
&  = & -R[\phi(\gamma_1)(\phi(\gamma_2)-R[\phi(\gamma_2)])], \eea
which evidently has only a first order pole, and that property
remains true for arbitrary products of primitives, and hence for
the whole Hopf algebra, if and only if $\phi$ is multiplicative.
Actually, most of the deficiencies of perturbation theory
vanish if we can evaluate with a $\phi$ which is a character
with respect to the product $\ast$, or $\vee$, for that matter.

The two crucial steps towards such a factorization, which amounts
to a partial resummation of graphs, are\\
- a requirement to absorb vertex subdivergences in Green functions
which depend only on a single scale, so that the beneficial
properties of one-parameter groups of scaling come to
bear, a requirement which sits very comfortably with the fact that gauge theories
relate vertex subdivergences to self-energies \cite{Slavnov},\\
- an appropriate use of the renormalization group in the
Dyson--Schwinger equations, which allows us to describe the presence
or absence of factorization in a controlled way in relation to the
fixpoint behaviour of the $\beta$-function of the theory.

That the renormalization group enters is quite obvious: the
structure of the Euler product as a product over geometric series
over residues of primitive graphs excludes any proliferation as
associated with a renormalon, a fact which by itself suggests that
if we are to achieve such a factorization, the renormalization
group should play a role. So these type of questions are certainly
of interest, and results along these lines will be pointed out in
upcoming work.

Finally, let us mention a first simple example as to how basic
algebraic structures of our graph insertions relate to physical
properties of a theory.

\begin{prop}
i) The product $\Gamma_1\vee\Gamma_2$ is integral for 1PI graphs
in $\phi^3_6$
and $\phi^4_4$.\\
ii) It is non integral for QED: $\Gamma_1\vee\Gamma_2=0\Rightarrow $
$ \Gamma_1=0$ or $\Gamma_2=0$ or $\Gamma_1=\Gamma_2=\vvp$.
\end{prop}
But now, the Hopf algebra of QED graphs can be divided by an
appropriate ideal of graphs $\Gamma$ containing $\vvp$ (the ideal
of graphs $\Gamma$ such that \ $\Delta^{bid(\Gamma)-1}(\Gamma)$ has
$\vvp$ as an element) and in the quotient -in which our product is
integral- it turns out that the Ward identities hold
automatically. The proposition has a generalization to non-abelian
gauge theories which is under scrutiny at the moment.

The final aspect in our outlook on QFT is about symmetries in the
Dyson--Schwinger equations which can relate them to differential
Galois groups. The equations are integral equations of a
complicated kind. But they still offer a lot of the symmetries
also known from differential equations. So a few short comments
along the lines of \cite{dennis} shall finish this section.
\subsection{Galois Groups and Feynman Graphs} There are many
symmetries in a Dyson--Schwinger equation, which reveal themselves
as invariants under the permutation of places where to insert
subgraphs, so they are reflected by identities between pole terms
of graphs. We have an obvious ring structure we are dealing with,
using products $\Gamma_1\vee\Gamma_2$ of 1PI graphs. We start
drifting towards a treatment of Feynman graphs as a ring, with
associated field of fractions say, where the role of primes is
played by primitive graphs, and an Euler product combined with an
appropriate shuffle identity for Feynman rules  should guide us
towards an appropriate notion of a $\zeta$-function for a given
Green function.  To get an idea what these symmetries are related
to, we remind ourselves that in the skeleton expansion of a
Dyson--Schwinger equation we sum over all possible insertion
places (gluing data). Indeed, the resulting series over graphs can
be written using elementary symmetric polynomials in the insertion
places, $\gamma^{[0]}$ say, of the skeleton $\gamma$.

So consider the combination $\Gamma_1(\ast_i-\ast_j)\Gamma_2$, the
{\em difference} of the insertion of a subgraph $\Gamma_2$ into
$\Gamma_1$ at two different places $i,j$.

Following \cite{InsElim,dennis} we can consider the "differential
equation" (here, {\small $Z_{[{\bf res}(\Gamma_2),\Gamma_2]}$
$(X)$} is a derivation which replaces $\Gamma_2$ by its tree-level
counterpart ${\bf res}(\Gamma_2)$ in $X$) \be Z_{[{\bf
res}(\Gamma_2),\Gamma_2]}(X)=\Gamma_1,\ee which is solved by the
bidegree two Hopf algebra element $X=\Gamma_1\ast_i\Gamma_2$ as
well as by the bidegree two $X= \Gamma_1 \ast_j\Gamma_2$.
Furthermore, the bidegree one primitive
$X=\Gamma_1(\ast_i-\ast_j)\Gamma_2$ solves the homogeneous equation
\be Z_{[{\bf res}(\Gamma_2),\Gamma_2]}(X)=0,\ee where we assume
throughout that $\Gamma_1$ and $\Gamma_2$ are of bidegree one. If
one linearizes a Dyson--Schwinger equation and restricts it to a
finite number of underlying skeletons, the equation, rewritten as
a differential equation, has many structural similarities with
differential equations which have regular singularities, as also
the above argument exemplifies. This suggests to connect the
insertion of subgraphs at various different places with Galois
symmetries, and is the motivation to indeed look at invariants
under such symmetries in Feynman graphs, with a beautiful first
result reported in \cite{BKK}: the coefficient of the highest
weight transcendental in the residues of two graphs connected by
such a symmetry is invariant. While this is obvious, thanks to the
scattering type formula, for the coefficient of the highest pole
in the regularization parameter, it is indeed a very subtle result
for the residue in a graph of large bidegree.
\subsection{Summary}
The interplay between number theory, noncommutative geometry and
perturbative quantum field theory reveals, to my mind, strong
hints towards the structure of quantum field theory. Many of the
ideas featured here are not to be brought to fruition quickly, but to my
mind it is a fascinating task for a theorist to unravel the
structures of the theories which have been most successful so far
in our description of nature, and which have been carefully
extracted from experimental evidence by the high energy and
condensed matter theoretical physics communities. The
combinatorial structures of renormalization with the relation to
the Riemann--Hilbert problem, the appearance of Euler--Zagier sums
as residues of diagrams, and the factorization properties of the
Dyson--Schwinger equations all point towards fundamental
mathematical structures. Recent ideas and progress in pure
mathematics \cite{Connes,Ramis} point towards quantum field
theory. We finally might get the message.
\section*{Acknowledgments}
It is a pleasure to thank Arthur Greenspoon for proofreading the manuscript.

\begin{verbatim}
CNRS-IHES, 35 Route de Chartres, F91440 Bures-sur-Yvette, France;
Center for Math.-Phys., Boston Univ., Boston MA02215, USA.
\end{verbatim}

\begin{thebibliography}{99}
\bibitem{Zinn-Justin} J.~Zinn-Justin, {\em Quantum Field Theory and
Critical Phenomena}, Oxford Univ.~Press (2002, 4th Ed.).
\bibitem{non-Aoki}
D.~Espriu, J.~Manzano and P.~Talavera, {\em Flavor mixing, gauge
invariance and wave-function renormalisation,} Phys.\ Rev.\ D {\bf
66}, 076002 (2002) [arXiv:hep-ph/0204085].
\bibitem{Marino}
M.~Marino, {\em Chern-Simons theory, matrix integrals, and
perturbative three-manifold  invariants,} [arXiv:hep-th/0207096].
\bibitem{diag}
G.~'t Hooft, M.~Veltman, {\em DIAGRAMMAR,}  Cern report 73/9
(1973), reprinted in "Particle interactions at very high
energies". NATO Adv. Study Inst. Series, Sect. B, vol. 4B, 177.
\bibitem{RHI}
A.~Connes, D.~Kreimer, {\em Renormalization in quantum field
theory and the Riemann-Hilbert  problem. I: The Hopf algebra
structure of graphs and the main theorem,} Commun.\ Math.\ Phys.\
{\bf 210} 249 (2000) [hep-th/9912092].
\bibitem{RHII}
A.~Connes, D.~Kreimer, {\em Renormalization in quantum field
theory and the Riemann-Hilbert  problem. II: The beta-function,
diffeomorphisms and the renormalization group,} Commun.\ Math.\
Phys.\ {\bf 216} 215 (2001)[hep-th/0003188].
\bibitem{InsElim}
A.~Connes and D.~Kreimer, {\em Insertion and elimination: The
doubly infinite Lie algebra of Feynman  graphs,} Annales Henri
Poincare {\bf 3}, 411 (2002) [arXiv:hep-th/0201157].
\bibitem{DK1}
D.~Kreimer, {\em On the Hopf algebra structure of perturbative
quantum field theories,} Adv.\ Theor.\ Math.\ Phys.\  {\bf 2}
 303 (1998)[q-alg/9707029].
\bibitem{Chen}
D.~Kreimer, {\em Chen's iterated integral represents the operator
product expansion,} Adv.\ Theor.\ Math.\ Phys.\  {\bf 3} 627
(2000) [hep-th/9901099].
\bibitem{dennis} D.~Kreimer, {\em Structures in Feynman graphs -
Hopf algebras and Symmetries}, talk given at the {\em Dennisfest},
Stony Brook June 14-21, 2001 [hep-th/0202110], to appear.
\bibitem{Book}
D.~Kreimer, {\em Knots and Feynman Diagrams}, Cambridge
Univ.~Press 2000.
\bibitem{BKold1}
D.~J.~Broadhurst, D.~Kreimer, {\em Knots and numbers in Phi**4
theory to 7 loops and beyond,} Int.\ J.\ Mod.\ Phys.\  {\bf C6}
 519 (1995)[hep-ph/9504352].
\bibitem{BKold2}
D.~J.~Broadhurst, D.~Kreimer, {\em Association of multiple zeta
values with positive knots via Feynman  diagrams up to 9 loops,}
Phys.\ Lett.\  {\bf B393}  403 (1997) [hep-th/9609128].
\bibitem{BGK}
D.~J.~Broadhurst, J.~A.~Gracey and D.~Kreimer, {\em Beyond the
triangle and uniqueness relations: Non-zeta counterterms at  large
N from positive knots,} Z.\ Phys.\ C {\bf 75} 559 (1997)
[arXiv:hep-th/9607174].
\bibitem{BK1}
D.~J.~Broadhurst, D.~Kreimer, {\em Renormalization automated by
Hopf algebra,} J.\ Symb.\ Comput.\  {\bf 27} 581 (1999)
[hep-th/9810087].
\bibitem{BK2}
D.~J.~Broadhurst, D.~Kreimer, {\em Combinatoric explosion of
renormalization tamed by Hopf algebra:  30-loop Pade-Borel
resummation,} Phys.\ Lett.\  {\bf B475} 63 (2000)
[hep-th/9912093].
\bibitem{BK4}
D.~J.~Broadhurst, D.~Kreimer, {\em Exact solutions of
Dyson-Schwinger equations for iterated one-loop  integrals and
propagator-coupling duality,} Nucl.\ Phys.\ B {\bf 600} 403 (2001)
[hep-th/0012146].
\bibitem{Goncharov}
A.B.~Goncharov, {\em Galois symmetries of fundamental groupoids
and noncommutative geometry}, math-ag/0208144.
\bibitem{review}
D.~Kreimer, {\em Combinatorics of (perturbative) quantum field
theory,} Phys.\ Reports {\bf 363} 387 (2002) [hep-th/0010059].
\bibitem{CK1}
A.~Connes, D.~Kreimer, {\em Hopf algebras, renormalization and
noncommutative geometry,} Commun.\ Math.\ Phys.\ {\bf 199} 203
(1998) [hep-th/9808042].
\bibitem{Kassel}
C.~Kassel, {\em Quantum Groups}, Springer 1995.
\bibitem{Loday}
J.-L.~Loday, M.O.~Ronco, {\em Trialgebras and families of
polytopes}, math-at/0205043, and references there.
\bibitem{Kurusch}
K.~Ebrahimi-Fard, {\em Loday-type algebras and the Rota-Baxter
relation}, Lett.\ Math.\ Phys.\ {\bf 61} 139 (2002)
[math-ph/0207043].
\bibitem{overl}
D.~Kreimer, {\em On overlapping divergences,} Commun.\ Math.\
Phys.\  {\bf 204} 669 (1999) [hep-th/9810022].
\bibitem{Fot}
F.~Markopoulou, {\em An algebraic approach to coarse graining},
hep-th/0006199.
\bibitem{BK}
D.~J.~Broadhurst, D.~Kreimer, {\em Towards cohomology of
renormalization: Bigrading the combinatorial Hopf  algebra of
rooted trees,} Commun.\ Math.\ Phys.\ {\bf 215} 217 (2000)
[hep-th/0001202].
\bibitem{Foissy}
L.~Foissy, {\em Les alg\`ebres des Hopf des arbres enracin\'es
d\'ecor\'ees}, Thesis, Univ.~Reims, Dept.~of Math., available from
the author: loic.foissy@univ-reims.fr (2001).
\bibitem{Connes}
A.~Connes, {\em Sym\'etries Galoisiennes et Renormalisation},
S\'em.\ Poincar\'e {\bf 2} 75 (2002) [math.qa/0211199].
\bibitem{CM}
A.~Connes, H.~Moscovici, {\em Hopf algebras, cyclic
co\-ho\-mo\-lo\-gy and the trans\-verse in\-dex theo\-rem},
Com\-mun.\ Math.\ Phys.\ {\bf 198} 199 (1998)  [math.\
dg/9806109].
\bibitem{Ramis}
J.-P.~Ramis, {\em Tr\`es anciennes et tr\`es nouvelles m\'ethodes
de sommation de s\'eries divergentes,} talk given at {\em
Colloque: Renormalization - Theory and Perspectives}, IHES,
Oct.14-18 2002.
\bibitem{Riv}
V.~Rivasseau, {\em An Introduction to Renormalization}, S\'em.\
Poincar\'e {\bf 2} 1 (2001).
\bibitem{Wight} A.S.~Wightman, {\em Some Lessons of Renormalization Theory},
in "The Lessons of Quantum Theory", J.~de Boer et.al.~Eds.,
Elsevier 1986.
\bibitem{KD}
D.~Kreimer, R.~Delbourgo, {\em Using the Hopf algebra structure of
QFT in calculations,} Phys.\ Rev.\  {\bf D60} 105025 (1999)
[hep-th/9903249].
\bibitem{brosnan}
P.~Belkale, P.~Brosnan, {\em Matroids, motives and a conjecture of
Kontsevich}, math.ag/0012198.
\bibitem{Rosner}
J.L.~Rosner, Phys.\ Rev.\ Lett.\ {\bf 17} 1190 (1966); Ann.\
Phys.\ {\bf 44}  11 (1967).
\bibitem{BDK}
D.J.~Broadhurst, R.~Delbourgo, D.~Kreimer, {\em Unknotting the
polarized vacuum of quenched QED}, Phys.\ Lett.\ {\bf B366} 421
(1996) [hep-ph/9509296].
\bibitem{Slavnov}
A.A.~Slavnov, {\em Regularization-Independent Gauge-Invariant
Renormalization of the Yang--Mills Theory}, Theor.\ Math.\ Phys.\
{\bf 130}(1) 1 (2002).
\bibitem{BKK}
I.~Bierenbaum, R.~Kreckel, D.~Kreimer, {\em On the invariance of
residues of Feynman graphs,}, J.\ Math.\ Phys.\ {\bf 43} 4721
(2002) [hep-th/0111192].
\end{thebibliography}
\end{document}